\documentclass[aps,prd,preprint,preprintnumbers,showpacs]{revtex4}
\usepackage{graphicx}

\begin{document}

\preprint{NSF-KITP-05-21, UCD0505}

\title{Latent heat of the large $N$ finite temperature phase transition}

\author{Joe Kiskis}
\email{jekiskis@ucdavis.edu}
\affiliation{Department of Physics, University of 
California\\ Davis, CA 95616, USA}

\date{\today}
       

\begin{abstract}

Reduced large $N$ gauge theories have a phase with unbroken center
symmetry and phases in which that symmetry is broken for Polyakov loops
in one or more lattice directions. The phase with unbroken symmetry is
associated with the zero temperature, infinite volume, infinite $N$
theory while the phase in which the symmetry is broken in just one
lattice direction has been conjectured to be the spatial reduction of
the high temperature, infinite volume, infinite $N$ theory. Measurements
of the scaling properties of the latent heat of the transition between
these phases test that hypothesis. The results indicate a non-zero
latent heat in the continuum limit. Substantial finite spacing effects
remain, and finer lattices will be needed to confirm physical scaling.

\end{abstract}


\maketitle

\section{Introduction}

Among the limited tools that can be used to investigate the
nonperturbative aspects of gauge theories are lattice numerical
calculations and large $N$ approximations. With continuing dramatic
advances in computer hardware, it is possible to combine the two and
make additional progress. Recent numerical results \cite{teper} 
\cite{teper_2} for $N$
up to 8 have confirmed that the finite $N$ corrections are surprisingly
small and that $N=3$ is remarkably close to $N=\infty$. However, since a
straightforward large $N$ simulation is more expensive than the physical
$N=3$ case, there is little motivation to follow that indirect route to
phenomenological results. The employment of reduction makes
large $N$ numerical results more interesting. 

Long ago, it was shown by Eguchi and Kawai \cite{ek} that certain
infinite volume, $N=\infty$ quantities (such as the free energy) can be
calculated in a reduced model where spacetime is reduced to one
point---provided the center symmetry $Z(N)$ of $SU(N)$ gauge theory is
unbroken.  Unfortunately long before the coupling $\lambda=Ng^2$ is
small enough to be near the continuum limit, the symmetry does break
\cite{neu}. Quenching \cite{neu} and twisting \cite{twist} were
developed as workarounds to this barrier. An alternative is to reduce
not from infinite volume to a single site lattice but to a lattice of
finite size $L^4$. For larger $L$, $\lambda$ can be pushed to smaller
values while remaining in the phase with unbroken $Z(N)$. An
investigation of this approach found evidence \cite{knn} that the
transition to the phase with the symmetry broken for Polyakov loops in
one lattice direction takes place at a physical value $L_c(\lambda) \sim
(a \Lambda_{QCD})^{-1}$. Thus the infinite volume, $N=\infty$ theory can
be simulated on a lattice of finite physical size.

However at each $L$, there is a critical coupling $\lambda_c$, which
depends on $L$, and for which the center symmetry is broken when
$\lambda<\lambda_c$. In \cite{knn}, it was observed that there is a
smaller coupling $\lambda_1$ defining a range $\lambda_1 < \lambda <
\lambda_c$ in which the $Z(N)$ symmetry is broken only for the Polyakov
loops in a single lattice direction. As $\lambda$ decreases further, the 
symmetry is broken in an increasing number of lattice directions. Does 
the phase with symmetry breaking in a single direction have physical
significance? In \cite{knn}, there is speculation that it is the large
$N$ limit of the finite temperature phase with $T > T_c$. The value of
$\lambda_c$ is roughly consistent with this claim, {\em i.e.}  $L_c
\approx 1/T_c$.

There is numerical evidence that the finite temperature phase
transition on large spatial lattices remains first order as $N$
increases \cite{teper} \cite{teper_2}. The latent heat $\Delta \epsilon$ 
appears to
approach the form $N^2 h$ with $h$ an $N$-independent physical energy
density.  In lattice units, $a^4 h$ scales as $(a \Lambda_{QCD})^4 \sim
L_c^{-4}$. Thus the latent heat should be accessible in the reduced
theory. The results in \cite{knn} indicated that the phase transition is
also first order in the reduced theory at large $N$.

If the phase of the reduced theory with the center symmetry broken in
one direction is indeed physical, our expectation is that measurements
of $a^4 h$ will have a non-zero limit for $N \rightarrow \infty$ and
$g_c^2 \rightarrow 0$ and will scale correctly with $L_c$ as $\lambda
\rightarrow 0$. This report presents measurements of $a^4 h$ on $L^4$
reduced lattices with $L=5, N=29; L=6, N=37; L=7, N=29$; and $L=8,
N=29$. Since the large $N$ corrections are O($1/N^2$), the expectation
is that they are negligible. Indeed some additional $L=6, N=29$ data
agree within statistical uncertainties with the $L=6, N=37$ result shown
below.  The results indicate that the continuum limit $\lambda
\rightarrow 0$ of $L_c^4 a^4 h$ is non-zero, but the asymptotic scaling
region in which it would be independent of $L_c$ has not been reached
for these $L_c$ values.

\section{Methods and results}

Numerical calculations follow those used in \cite{knn}. The Monte Carlo
evolution uses the standard Wilson gauge action. Each update of the
lattice consisted of a heatbath sweep followed by a overrelaxation
sweep. In the heatbath sweep, the SU(N) group element on a link is
modified by working with each of its $N(N-1)/2$ SU(2) subgroups in
turn. A typical run was for one or two thousand lattice updates.

The phase is determined by monitoring the four quantities
\begin{equation}
P_{\mu}=\frac{1}{N^2} \langle \sum_{i,j}^N 
\sin^2[\frac{1}{2}(\theta_i-\theta_j)] \rangle .
\end{equation}
The $\theta's$ are the angles of the eigenvalues of a Polyakov
loop in the $\mu$ direction. The average is over sites in a plane
perpendicular to that direction as described in \cite{knn}. For unbroken 
symmetry, $P_{\mu}=0.5$, and it decreases for symmetry breaking in the 
$\mu$ direction. 

By varying the coupling at fixed $L$, one can identify the narrow region
of metastability where both the symmetric phase and the phase with the
symmetry broken for one lattice direction can exist for substantial
periods of Monte Carlo time.  In terms of $b =1/(Ng^2) =1/\lambda$, the
critical values are at about 0.347, 0.352, 0.356, and 0.3595 for $L=$ 5,
6, 7, and 8 respectively. The metastable regions are of width about
0.0005 in $b$, {\em i.e.} about a part in a thousand. At each $L$, a
value for the jump in the average plaquette $\Delta s$ is obtained at a
$b$ where one phase is stable and the other metastable. For $L=8$, this
could be done at one $b$ value, while for the other $L$s, two $b$s
separated by 0.0005 were possible. 

Table~\ref{tab:table1} shows the results for $L_c^4 \Delta s$ The
normalization of the average plaquette is such that it approaches one as
the coupling goes to zero. In the cases where there are two $b$ values
at a single $L$, there is also a line that shows the average of the two
jumps. The uncertainty in the jump at one $b$ value is based on the
statistics of several runs of one or two thousand sweeps at that $b$
value. The differences in the jumps for two $b$ values at the same $L$
are larger than the uncertainties in each jump.  Thus there is a
systematic uncertainty associated with the choice of $b$. The uncertainty
in an average jump is estimated as the difference between the jump values
at the two $b$ values.


\begin{table}
\caption{\label{tab:table1} Data for $L_c^4 \Delta s$.}
\begin{ruledtabular}
\begin{tabular}{ccc}
L&b& $L^4 \Delta s$ \\
\hline
5 & 0.3475 & 1.456 $\pm 0.03$\\
5 & 0.3470 & 1.584 $\pm 0.04$\\
5 & average & 1.520 $\pm 0.064$\\
6 & 0.3520 & 1.199 $\pm 0.026$\\
6 & 0.3515 & 1.037 $\pm 0.04$\\
6 & average & 1.118 $\pm 0.08$\\
7 & 0.3560 & 0.816 $\pm 0.05$\\
7 & 0.3555 & 1.0084 $\pm 0.05$\\
7 & average & 0.9124 $\pm 0.096$\\
8 & 0.3595 & 0.778 $\pm 0.06$\\
\end{tabular}
\end{ruledtabular}
\end{table}


The jump in the average plaquette $\Delta s$ is proportional to 
the discontinuity in $\epsilon-3P$ \cite{engels} \cite{text}. Since
the pressure $P$ is continuous at a transition, the jump in the
plaquette also gives the jump in the energy density $\epsilon$
and thus $a^4h$.

\begin{figure}
\centering
\includegraphics[scale=0.5]{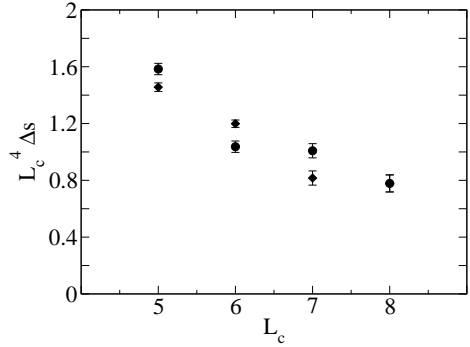}
\caption{\label{fig1}The values of $L_c^4\Delta s$ from 
Table~\ref{tab:table1} with definite $b$ are shown.}
\end{figure}

Figure~\ref{fig1} uses the data from Table~\ref{tab:table1} at definite
$b$ values and plots $L_c^4 \Delta s$ as a function of $L_c$. This would
be a constant in the continuum limit, but there are substantial
variations from that limiting case. Most likely these are a finite
spacing effect, {\em i.e.} the coupling is not sufficiently small or
$L_c$ sufficiently large to see continuum scaling. For comparison,
consider the results \cite{teper_2} or \cite{gavai} in non-reduced
calculations.  At $N_t$ values that are the same as the $L_c$ values
used here, the scaling violations are also substantial and comparable to
results shown here.

\begin{figure}
\centering
\includegraphics[scale=0.5]{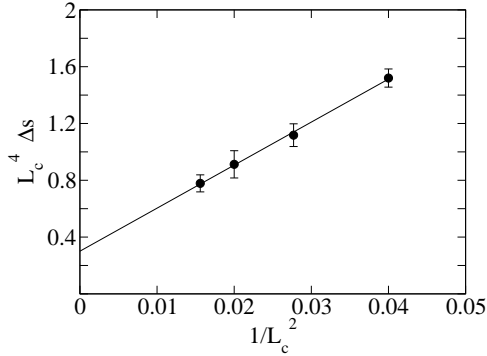}
\caption{\label{fig2}The average values for $L_c^4 \Delta s$ 
from Table~\ref{tab:table1} are plotted along with the line that gives 
the best fit.}
\end{figure}

If the dominant scaling violations at these lattice spacings are an 
$a^2$ correction, then $L^4 
\Delta s$ {\em vs.} $L^{-2}$ would be a line. 
Figure~\ref{fig2} shows the average data plotted in that way. A straight
line fit to the data has a $L^4 \Delta s$ intercept of 0.30. On the
other hand, if the $L=8$ point were run in assuming a perturbative
$\beta$ function, the intercept would be considerably higher. 

Thus it appears that these calculations are not at $L_c$ values that are
sufficiently large ($\lambda$ sufficiently small) to confirm weak
coupling continuum scaling for a physical value of the
latent heat. However, since the results for the jump in the plaquette
are close in value to the non-reduced results and show a similar scaling
violation, the hypothesis that the phase of the reduced model with the
center symmetry broken in one direction is the reduced $T>T_c$ phase
remains viable. Pushing the calculations to larger $L$ would be fairly
expensive. 

To assess the strength of the transition, it is conventional to compare 
the latent heat per unit three volume in lattice units 
\begin{equation}
 a^4 \Delta \epsilon = N^2 a^4h = -12 N^2 a \frac{\partial}{\partial a} 
             \frac{1}{\lambda} \Delta s
\end{equation}
with the blackbody energy density per massless vector particle
\begin{equation}
 \epsilon_{SB} = 4 \sigma T^4
\end{equation}
with
\begin{equation}
 \sigma = \frac{\pi^2}{60}
\end{equation}
the dimensionless Stephan-Boltzmann constant.
In the continuum limit,
\begin{equation}
  -a\frac{\partial}{\partial a} \frac{1}{\lambda} = \frac{11}{24 \pi^2}.
\end{equation}
\begin{equation}
 \frac{\Delta \epsilon}{N^2 \epsilon_{SB}} = \frac{165}{2 \pi^4} L^4 
\Delta s
\end{equation}
For an $L^4 \Delta s$ intercept of 0.30, this is 0.26.

If the phase with intact center symmetry is the reduced zero temperature 
theory, and the phase with center symmetry broken in one direction is 
the reduced $T > T_c$ phase, one may wonder what has happened to the 
phase with $ 0<T<T_c$. In fact, the finite temperature effects in this 
region are nonleading in $N$ and are therefore invisible in the reduced 
theory which captures only the leading $N^2$ term correctly 
\cite{gn} \cite{cohen} \cite{thorn}.

\section{Conclusion}

Reduced large $N$ gauge theories have a transition between phases with
unbroken and broken center symmetry. If the phase with the center
symmetry broken in one direction is indeed physical, our expectation is
that $a^4 h$ will have a non-zero limit for $N \rightarrow \infty$ and
$g_c^2 \rightarrow 0$ at fixed $\lambda=Ng^2$ and will scale correctly 
with $L_c$ as $\lambda \rightarrow 0$. The results
show that $a^4 h$ is non-zero in the limit $N \rightarrow \infty, g^2
\rightarrow 0$ with $\lambda$ fixed and also indicate that $L_c^4 a^4 h$
has a non-zero limit as $\lambda \rightarrow 0$. However, substantial
finite spacing effects remain, and finer lattices will be needed to
confirm asymptotic physical scaling. This means that $h$ has a finite,
non-zero value in physical units, but the proper functional form for its
approach to the continuum limit remains to be confirmed.

\begin{acknowledgments}

I am grateful to Rajamani Narayanan and Herbert Neuberger for insightful 
discussion, valuable suggestions, and for commenting on this manuscript.

This research was supported in part by the National Science Foundation
under Grant No. PHY99-07949.

\end{acknowledgments}

\bibliography{paper_spires}

\end{document}